\documentclass[aps,twocolumn,noshowpacs,noshowkeys,superscriptaddress
]{revtex4}
\usepackage{amsfonts}
\usepackage{amsmath}
\usepackage{amssymb}
\usepackage{graphicx}

\usepackage{textcomp}%
\begin{document}

\def\neel{Institut N\'{e}el, CNRS et Universit\'{e} Joseph Fourier, 38042 Grenoble, France}
\def\cotton{
Laboratoire Aim\'e Cotton, CNRS, Universit\'e Paris-Sud and ENS Cachan, 91405 Orsay, France}

\author{S.~Rohr}
\affiliation{\neel}
\author{E. Dupont-Ferrier}
\affiliation{\neel}
\author{B. Pigeau}
\affiliation{\neel}
\author{P. Verlot}
\affiliation{\neel}
\author{V. Jacques}
\affiliation{\cotton}
\author{O. Arcizet}
\affiliation{\neel}
\email{olivier.arcizet@grenoble.cnrs.fr}
\title{Synchronizing the dynamics of a single NV spin qubit on a parametrically coupled radio-frequency field through microwave dressing}

\begin{abstract}
A hybrid spin-oscillator system in parametric interaction is experimentally emulated using a single NV spin qubit immersed in a radio frequency (RF) field and probed with a quasi resonant microwave (MW) field. We report on the MW mediated locking of the NV spin dynamics onto the RF field, appearing when the MW driven Rabi precession frequency approaches the RF frequency and for sufficiently large RF amplitudes. These signatures are analog to a phononic Mollow triplet in the MW rotating frame for the parametric interaction and promise to have impact in spin-dependent force detection strategies.
\end{abstract}
\maketitle

In recent years, research on mechanical hybrid systems has powered vast experimental and theoretical efforts with the counter intuitive perspectives of revealing the quantum behavior of matter at the macroscopic scale \cite{Schwab2005,Treutlein2013}, such as the creation of non-classical states of motion. These semi-quantum devices generally consist of a mechanical oscillator interfaced with a quantum system, such as circuit qubits \cite{LaHaye2009,Pirkkalainen2013},  cold atoms \cite{Treutlein2007}, quantum dots \cite{Lassagne2009, Steele2009, Sallen2009, Bennett2010} or single molecules \cite{Ganzhorn2013}, with which quantum state swapping is envisioned as in pioneering trapped ion experiments \cite{Blatt2008}. Due to their unique spin coherence properties \cite{Hanson2006, Childress2006, Balasubramanian2009} which potentially allow to enter the strong coupling regime \cite{Rabl2009,Rabl2010}, NV defects have received important attention. In particular the influence of mechanical motion on the spin dynamics has recently been investigated \cite{Arcizet2011a, Bennett2012, Yacoby2012}, while demonstrating the reverse interaction remains yet a challenging task.\\
Many of the advanced protocols envisioned for observing spin dependent forces \cite{Rugar2004, Degen2009, Rabl2010, Nichol2012} -the key ingredient for nonclassical state generation- would benefit from the ability to manipulate the spin state at the mechanical oscillation frequency, permitting an enhancement of the spin dependent force readout capacity.
This represents an experimental challenge since the mechanical resonances are designed to be spectrally sharp in order to gain access to large force sensitivities enabled by high mechanical quality factors while the experimental control that can be obtained on the stability of a single spin precession frequency is significantly lower.\\
In this letter we emulate a hybrid spin-mechanical system by immersing a single NV spin in an external radio-frequency (RF) field which modulates the qubit energy \cite{Childress2010,Li2013} in analogy to a mechanical oscillator in parametric interaction with the two level system (TLS). We report on the observation and analysis of synchronization of the single spin dynamics on the emulated mechanical oscillator motion when driven at sufficiently large amplitudes, while adjusting the MW power to make the spin precession approach the oscillation frequency. In this regime, the RF/mechanical modulation is dynamically imprinted on the qubit dynamics, allowing an enhanced coherent spin backaction onto the mechanical oscillator once implemented in hybrid mechanical systems.
\begin{figure}[b]
\begin{center}
\includegraphics[width=\linewidth]{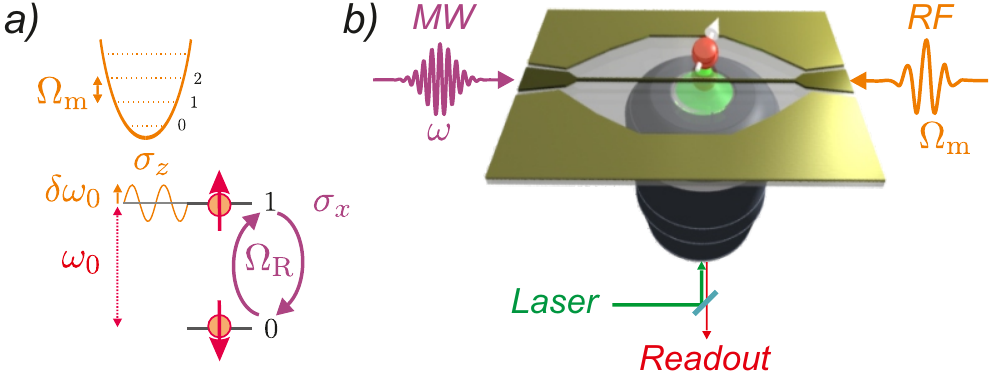}
\caption{An emulated spin-mechanical hybrid system. a) A single NV spin qubit is coupled to a RF field (frequency $\Omega_{\rm m}/2\pi$) which parametrically modulates its energy splitting  ($\hbar\omega_0$) with an amplitude $\hbar\delta\omega_0$ while  quasi-resonant MW irradiation (pulsation $\omega$) is used to manipulate the spin state, drive Rabi precessions of the TLS (frequency $\Omega_R/2\pi$) and realize resonant spectroscopy of the system. b) Experimental setup:  a single NV defect hosted in a diamond nanocrystal deposited on a glass plate is optically readout through a confocal microscope apparatus and the spin state dependent fluorescence  \cite{Jelezko2004} is readout on an avalanche photodiode. MW and RF fields required for spin manipulation \cite{Jelezko2004} and spin energy modulation, respectively, are delivered via a coplanar waveguide (see SI).}\label{fig1}
\end{center}
\end{figure}
Most of qubit-based sensing protocols, originally developed in the field of NMR on spin ensembles \cite{Slichter1996}, rely on dynamical manipulation of the spin state at the frequency of the signal under investigation by combining perpendicular signal and control fields \cite{Holmstrom1997, Holmstrom1998, Saiko2008, Saiko2010}. These techniques have been implemented in qubit environment spectroscopy or ultra-sensitive sensing experiments \cite{Kotler2011,Lange2011,Degen2013} aiming at detecting weak signals for which the spin synchronization is not visible.
The MW dressing enables the original parametric phonon coupling to be turned into a resonant interaction -the one of cavity QED \cite{Haroche2006}- where a phononic excitation generates transitions between dressed states. Since the traditional roles of $\sigma_x$ and $\sigma_z$ operators are interchanged in the picture of dressed states, the synchronization can be understood as a Mollow triplet signature of the parametric interaction.

\textit{Formalization---}A qubit parametrically coupled to a RF field- or similarly a hybrid spin-mechanical system- is described by the Hamiltonian \cite{Rabl2009,Treutlein2013} $H=H_0+H_{\rm int}$ with $H_0=\hbar \omega_0\sigma_z+\hbar \Omega_{\rm m} a^\dagger a +\hbar \omega b^\dagger b $ where $\hbar \omega_0$ is the qubit energy splitting, $\Omega_{\rm m}/2\pi$ the oscillator/RF frequency, and $\omega/2\pi$ the frequency of the MW  field used for spin manipulation. The $a, a^\dagger$ and $b, b^\dagger$ operators are annihilation and creation operators for the RF and MW fields respectively. The interaction Hamiltonian is $H_{\rm int}=\hbar \kappa^v (a+a^\dagger)\sigma_z+\hbar \Omega_R^v \sigma_x (b+b^\dagger).$
The first term describes the parametric interaction with the RF-oscillator field, $\kappa^v$ being the vacuum coupling strength. For a classically driven phonon field  it can be rewritten as $\hbar \sigma_z\, \delta\omega_0 \cos(\Omega_{\rm m}t)$,  $\delta\omega_0$ describing the amplitude of the parametric energy modulation (see Fig. 1a).
The second term represents the MW field interaction with the spin. For a classical MW field, it can be similarly replaced by $\hbar \Omega_R \sigma_x\cos(\omega t)$ where $\Omega_R$ represents the strength of the MW field, with a detuning defined by $\delta\equiv\omega-\omega_0$. Although the quantization of the phonon and photon fields is not formally necessary to describe our experiment where the only quantum object is the NV spin qubit, it permits for an elegant description of our findings in analogy with the appearance of Mollow triplets in dressed states spectroscopy.\\

\begin{figure}[t]
\begin{center}
\includegraphics[width=\linewidth]{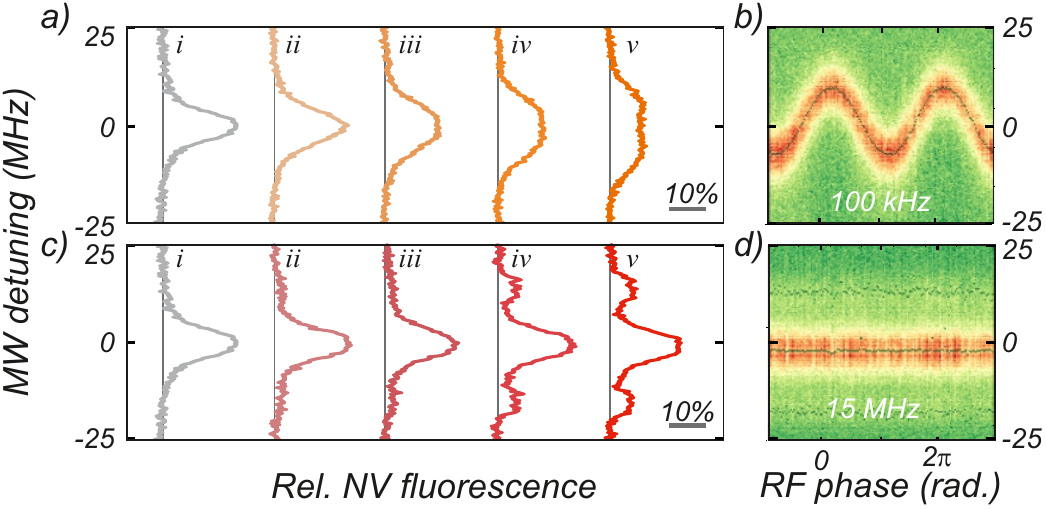}
\caption{Characterization of the spin-RF interaction in the adiabatic and resolved sideband regimes ($\Omega_{\rm m}/2\pi = 20\,\rm kHz$ in (a,b),  $15\,\rm MHz$ in (c,d)).  Left: ESR spectra obtained in absence of RF field ({\it i}) and for increasing oscillation amplitudes (resp. $\delta\omega_0 =  2.5, 4.7, 6.2\,  \rm and\, 9.1 \,\rm MHz$ from {\it ii} to {\it v}). Right: Time resolved ESR spectra obtained by gating the photon counts ($20\,\%$ duty cycle) at different RF oscillatory phases for $\delta\omega_0=9.1\,\rm MHz$. In the adiabatic case (b), the modulation observed has an amplitude of $\delta\omega_0$, while no temporal evolution is visible in the RSB case (d) since the spin dynamics is too slow to follow the imposed RF field.}\label{fig2}
\end{center}
\end{figure}
\textit{Spin-RF interaction---} The parametric interaction between the RF field and the NV spin is first characterized by continuous electron spin resonance (ESR) measurements (see Fig. 2), where the fluorescence of the NV defect is recorded as a function of the applied MW frequency. In absence of RF field (Fig. 2.a.{\it i} and 2.c.{\it i}) a characteristic fluorescence reduction is observed when the MW frequency is resonant with the spin transition. The RF field amplitude is subsequently progressively increased ({\it i} to {\it v}). When the mechanical frequency is small enough compared to the internal spin dynamics (a: $\Omega_{\rm m}/2\pi=20\,\rm kHz$), we observe a motional broadening of the spin resonance, as observed in earlier hybrid qubit-mechanical experiments \cite{Arcizet2011a,Kolkowitz2012,Yacoby2012,Li2013}. RF phase-gated ESR measurements synchronized on the RF modulation phase enable the observation of  the  spin resonance temporal evolution and measuring the  energy modulation induced by the RF amplitude $\delta\omega_0$ (Fig. 2b). In contrast, when the RF frequency is larger than the spin decay rate, sidebands appear on the ESR spectra. As a hallmark of frequency modulation, they are characterized by a separation corresponding to the RF oscillation frequency (15 MHz here) and a depth which initially increases with the oscillation amplitude (Fig. 2c, 2d). The strength of the sidebands \cite{Childress2010,Li2013,Pirkkalainen2013,Puller2013,Oliver2005,DupontFerrier2013}, defined as their relative area in order to account for MW and optical broadening \cite{Dreau2011}, has been experimentally verified to follow a Bessel evolution and is in agreement with our numerical simulations of Bloch equations (see SI). The substructure especially visible in the sidebands reflects the hyperfine coupling between  the NV electronic and $\rm ^{14} N$ nuclear spins. For the remainder of this letter all considerations will be restricted to the resolved sideband regime, where the mechanical oscillation frequency is larger than the spin decay rates, which have been measured at the level of $T_1\approx 173\, \rm{\mu s}$ and  $\Gamma_{\rm spin}=3\times10^5 \,\rm s^{-1}$ ($T_2^{\rm Rabi}\approx 3\,\rm \mu s$) at $\Omega_R/2\pi=5 \,\rm MHz$ (see SI).\\

\begin{figure}[b]
\begin{center}
\includegraphics[width=\linewidth]{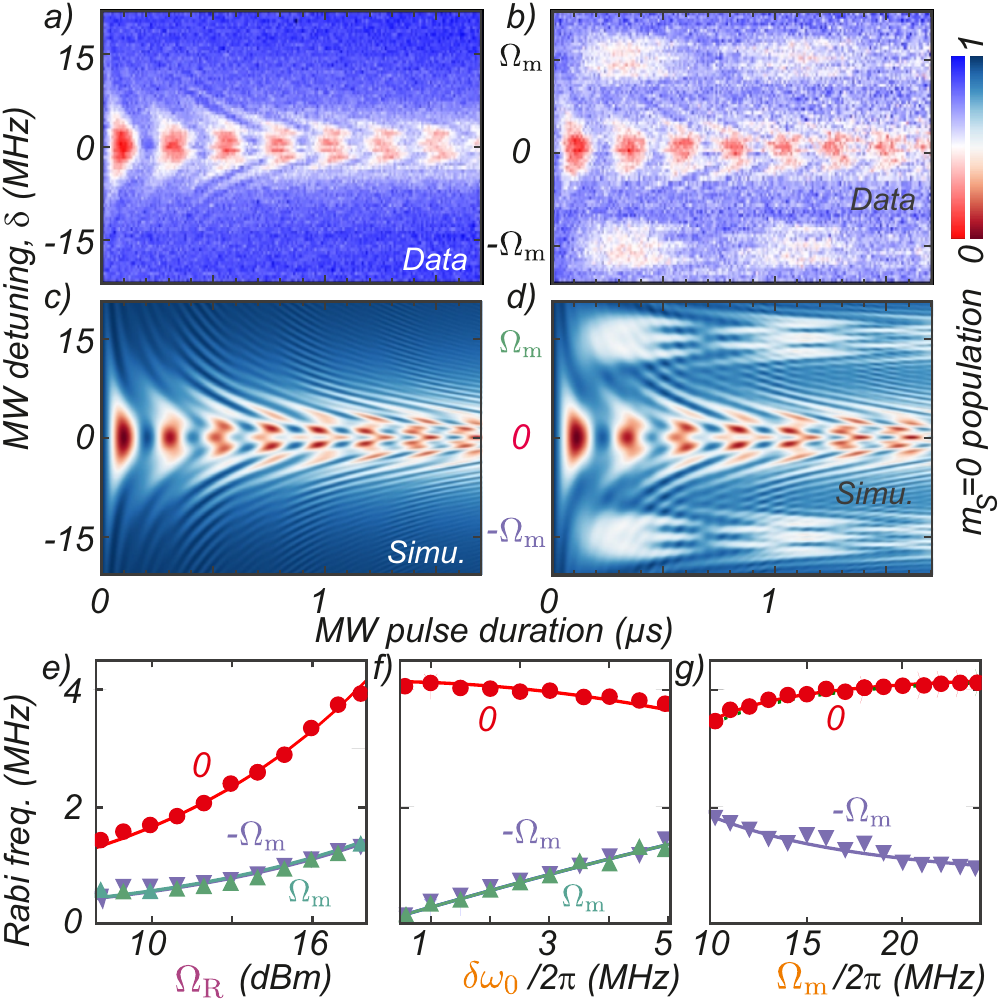}
\caption{Experimental (a,b) and numerically simulated (c,d) Rabi oscillations of NV spin in presence (b,d) and absence (a,c) of RF modulation. The spin state is optically prepared in its $m_S=0$ ground state and subsequently irradiated with a MW field whose detuning is varied across the resolved  sideband spectrum of Fig. 1c. The substructures observed at large MW pulse duration are inherent to the NV defect's hyperfine structure. Bottom: evolution of the Rabi frequency measured by tuning the MW frequency on the carrier ($\delta=0$) or on the first sidebands ($\delta=\pm\Omega_{\rm m}$) as a function of the driving amplitude $\Omega_R$ (e), oscillation amplitude $\delta\omega_0$(f) and oscillation frequency ($\Omega_{\rm m}/2\pi$) (g). When non varied, parameters are $\delta\omega_0/2\pi=4.5 \,\rm MHz$, $\Omega_{\rm m}/2\pi=15\,\rm MHz$ and 18\,dBm MW power.   Solid lines are fits based on Bessel functions and calibration measurements, see text.}
\end{center}
\end{figure}

\textit{Rabi maps---}In order to quantify the spin-MW interaction, we have carried out a systematic study of MW induced Rabi oscillations of the NV qubit. The measured Rabi oscillations are shown in Fig. 3 in absence (a) and presence (b) of RF parametric drive, and the corresponding numerical simulations can be seen in panels (c,d). The measurements performed on the non-RF driven NV qubit are used to calibrate the Rabi frequency which presents an initial quadratic dependence in the MW detuning. When the RF drive is turned on, the structure of the Rabi map is significantly modified. Rabi oscillations can now be driven when pumping the spin on the motional sidebands by adjusting the MW detuning to $\delta=\pm\Omega_{\rm m}$.
The Rabi oscillation frequency $\Omega_R^n\, (n=0,\pm1)$ has subsequently been measured for varying MW powers (e), RF amplitudes (f) and RF frequencies (g) and compared to the equation $\Omega_R^n=\Omega_R\left|J_n(\delta\omega_0/\Omega_{\rm m})\right|$, where $J_n(x)$ is the $\rm n^{th}$-order Bessel function of the first kind. Note that these measurements have been carried out at sufficiently low MW powers so that the Rabi precession frequency remains small compared to the RF frequency. This capacity of driving coherent spin oscillations on the motional sidebands is of importance in hybrid spin-oscillator systems since the MW photon absorption simultaneously proceeds through phonon emission/absorption ($\delta = \pm\Omega_{\rm m}$), which enables advanced qubit-based cooling protocols \cite{Rabl2010, Pirkkalainen2013} in analogy with laser cooling of ions \cite{Leibfried2003} or optomechanical cooling \cite{Schliesser2009} experiments. In the following, the experiments are carried out at resonance ($\delta=0$) and at the excited state level anti-crossing \cite{Jacques2009,Smeltzer2009} where electron-nuclear spins flip-flops mediate a $^{14}N$ nuclear spin polarization, liberating our measurements from beating frequencies due to the hyperfine coupling.

\begin{figure*}[t!]
\begin{center}
\includegraphics[width= 0.9 \linewidth]{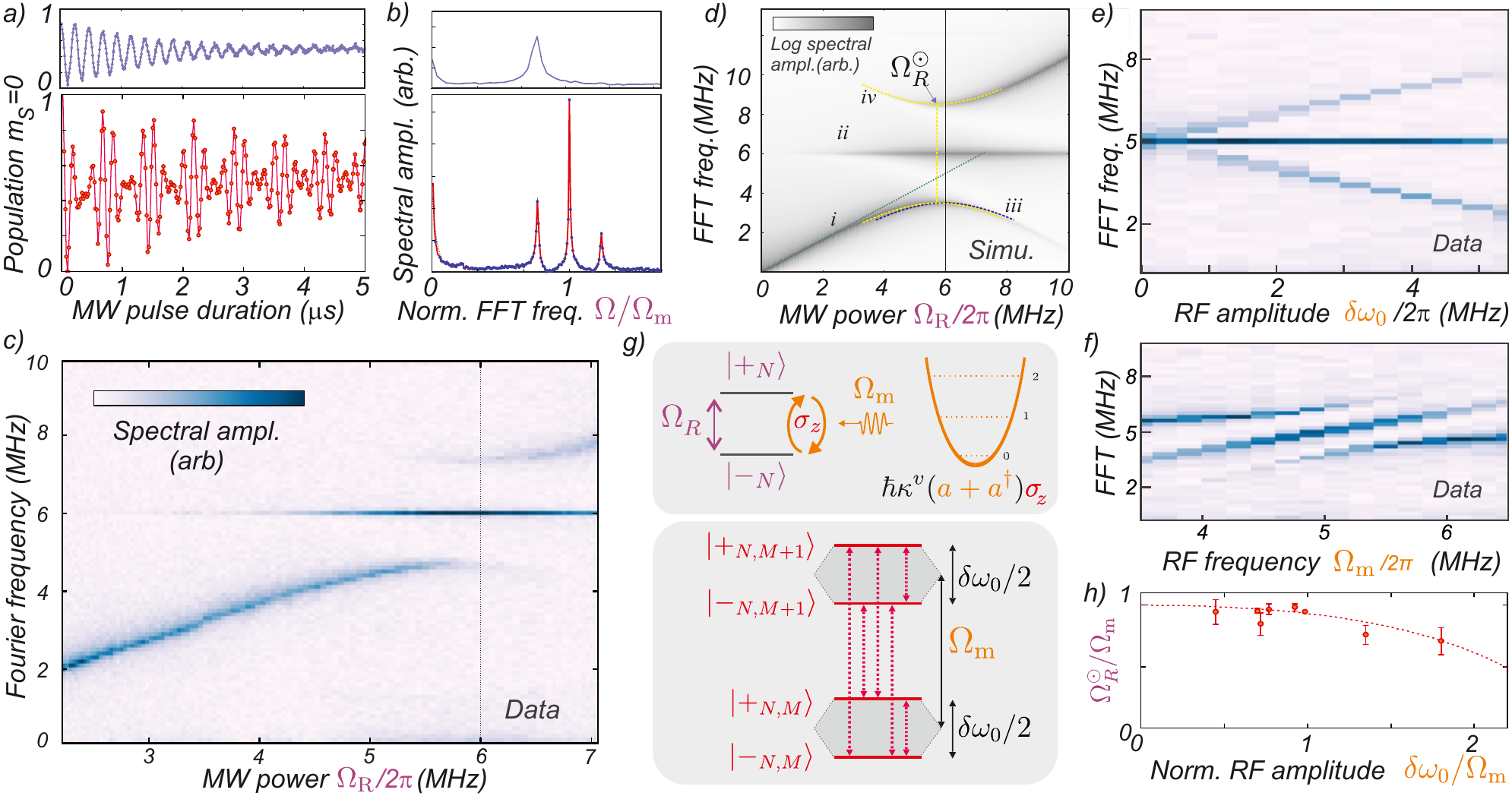}
\caption{Spin synchronization. Rabi oscillations (a) and their corresponding FFT spectra (b) measured for resonant MW pumping ($\delta=0$) in absence (above) or presence (below) of RF modulation at a frequency close to the Rabi precession: $\Omega_R\approx\Omega_{\rm m}$ (5/6 MHz here).  The triplet structure is the signature of spin locking, which is more easily evidenced in  (c,d): experimental and simulated FFT spectra of the Rabi oscillations for varying MW power to scan the Rabi precession frequency across the RF/mechanical resonance. In the spin locking region, a gap appears in the energy spectrum of spin dynamics. (e): FFT spectra measured at the synchronization point ($\Omega_{\rm R} = \Omega_{\rm m}$) for increasing RF amplitudes $\delta\omega_0$. (f): variation of the mechanical frequency across the synchronization point for fixed MW power ($\Omega_R/2\pi=5\,\rm MHz$) and RF amplitude $\delta\omega_0/2\pi = 2.7\,\rm MHz$. (g): Interpretation of the synchronization mechanism in term of RF dressing of the MW dressed qubit. Transitions between doubly dressed states are detectable as frequency components of Rabi oscillations: $\Omega_{\rm m},\Omega_{\rm m}\pm\delta\omega_0/2$.(h):  "RF light shifts": dependence on the oscillation amplitude of MW powers $\Omega_R^\odot$ which minimizes the triplet splitting. The dashed lines represent 6th order expansions of the Bloch-Siegert light shifts (see text), while data points are derived from 8 measurements similar to panel c.}
\end{center}
\end{figure*}

\textit{Spin synchronization---} The spin synchronization mechanism appears when the MW amplitude is chosen so that the Rabi precession approaches the RF/mechanical oscillation frequency: $\Omega_R\approx\Omega_{\rm m}$. In that case, and for sufficiently large RF amplitudes ($\delta\omega_0\gtrsim\Gamma_{\rm spin}$), the Rabi dynamics of the NV spin qubit is dramatically modified, as shown in Fig. 4a and in the corresponding Fourier analysis (Fig 4b) where a triplet structure is observed.
The synchronization of the spin precession frequency onto the RF drive frequency can be better visualized when the MW power is scanned so that the Rabi frequency is swept across the synchronization region (Fig. 4c). When the Rabi frequency is sufficiently detuned from the mechanical oscillation frequency, only one single peak is visible in the FFT, corresponding to the regime explored above (Fig 3.). When increasing the MW power to approach the synchronization region, a triplet structure appears with a dominant central component oscillating exactly at the RF frequency and a minimum half splitting of $\delta\omega_0/2$. The impossibility of generating Rabi oscillations  at frequencies lying in the gaps to both sides of the central peak illustrates the spin synchronization onto the RF-mechanical field. The MW pulse sequence is intentionally not synchronized to the RF phase in these experiments. We further investigated this mechanism by varying the RF oscillation amplitude $\delta\omega_0$. As can be seen on Fig. 4e, the triplet separation - that is the synchronization capture region - increases linearly with the RF drive. Note that the spin locking mechanism \cite{Slichter1996} which has recently been implemented on NV and trapped ions  magnetometers \cite{Degen2013,Kotler2011}, is usually investigated at low RF drive amplitudes, for which spin synchronization as shown here can not be observed. Our experimental findings are in quantitative agreement with numerical simulations based on Bloch equations (Fig. 4d)(see SI), except for peak linewidth. This is a consequence of the well established non-Markovian nature of the spin bath in diamond \cite{Lange2011}, which is also responsible for the observed extended Rabi decay times (Fig. 4a).\\
The synchronization mechanism can be qualitatively understood through a parametric driving of the spin precession, a virtual oscillator whose frequency depends quadratically on the MW detuning: $\Omega_R^{\rm eff}\approx\Omega_R+\delta^2/2\Omega_R$ (see Fig. 3). The RF field modulates the detuning  at $\Omega_{\rm m}$ and thus the effective Rabi frequency at $2\Omega_{\rm m}$, which falls close to twice the precession frequency in the synchronization region where $\Omega_R\approx\Omega_{\rm m}$. This regime of parametric driving of the Rabi precession generates an amplified spin precession at half the parametric modulation frequency, $\Omega_{\rm m}$, which is then locked onto the RF field (see Fig. 4.f) with a capture frequency range increasing with the drive amplitude $\delta\omega_0$ (see Fig 4.e) as in classical parametric driving of oscillators. This interpretation is of course limited, since the complex qubit dynamics on the Bloch sphere can not be fully reproduced in this simplified description.

\textit{Doubly dressed qubit interpretation---} The spin locking mechanism can be better interpreted by considering the qubit as being doubly dressed with MW and RF fields. We first proceed with a MW-dressing of the spin qubit \cite{Haroche2006}, resulting in a new energy ladder of dressed states $|\pm_N\rangle$, of energy splitting $\Omega_R^v\sqrt{N}$ increasing with the parameterized photon number $N$. In case of large coherent MW drive, the study can be restricted to the dressed states associated with the coherent state mean photon number whose splitting amounts to $\Omega_R$ (see SI). This direct manifestation of the MW dressing represents a new effective two level system (see Fig. 4g), at the origin of the Autler-Townes doublet \cite{Autler1955}, between which the $\sigma_z$ operator (the one precisely involved in the hybrid coupling hamiltonian) has non zero matrix elements. The synchronization mechanism then appears as a consequence of  RF/mechanical phonon parametric interaction with the MW-dressed qubit, which permits a second RF dressing step. This generates a new set of multiplicities that are now parameterized by the phonon number $M$  inside of which the new eigenstates $|\pm_{N,M}\rangle$ are linear combinations of product states $|\pm_N\rangle|M\rangle$ (see Fig 4g). For strong coherent RF fields, the energy splitting can be expressed as $\Delta=\sqrt{(\Omega_{R}-\Omega_{\rm m})^2+\delta\omega_0^2/4}$. The experimentally measured FFT spectra of Rabi oscillations can be obtained by calculating the spectrum of the $\sigma_z$ operator within the doubly dressed states. In analogy with the Mollow triplet in quantum electrodynamics, the Rabi oscillations spectrum will contain signals at the mechanical frequency $\Omega_{\rm m}$ but also sidebands at $\Omega_{\rm m}\pm\Delta$, as indeed observed in the experimental spectra of Fig. 4b. Note that the Mollow triplet structure is only visible for large RF/mechanical fields, for which $\delta\omega_0/2>\Gamma_{\rm spin}$. Since no synchronization is required for observing the triplet structure, these signatures are expected to persist with mechanical motion of reduced temporal coherence such as Brownian motion or zero-point fluctuations for sufficiently high coupling strength.

\textit{RF light shifts---} To fully explain the experimental observations, the rotating wave approximation that is implicitly done in the dressing procedure is not entirely justified since the RF driving amplitudes $\delta\omega_0$ can be of equal magnitude as the mechanical frequency $\Omega_{\rm m}$ (corresponding to the ultra-strong coupling regime). Indeed one can clearly see from the experimental data (Fig 4c), but also from numerical simulations (Fig. 4d) that the MW power $\Omega_R^\odot$ for which the triplet structure presents the smallest energy splitting is slightly lower than $\Omega_{\rm m}$. This is a direct signature of the "light shifts" or Bloch Siegert effect \cite{Cohen-Tannoudji1973,Saiko2008} of the RF field on the MW-dressed spin qubit. We have carried out a numerical and experimental analysis of the RF light shifts through the dependence of $\Omega_R^\odot$ in the RF oscillation strength (see Fig. 4.h  and SI),which are in good agreement with the analytical expansion in $\delta\omega_0/\Omega_{\rm m}$ of QED light shifts \cite{Cohen-Tannoudji1973}.\\
In conclusion, the synchronization criteria $\Omega_{\rm m} \approx \Omega_R$ and $\Omega_{\rm m}, \delta\omega_0/2 > \Gamma_{\rm spin}$ can be achieved in state of the art implementations. For example, a $50\times 0.05\,\rm\mu m$ SiC nanowire oscillating at $1\,\rm MHz$ \cite{Nichol2012,Arcizet2011a,Rabl2009} with a 3\,nm rms amplitude at 300\,K and a 700\,fm zero point motion allows achieving coupling strengths ($\delta\omega_0, \kappa^v$) of (80\,MHz, 20\,kHz) respectively once immersed in strong magnetic field gradients of $10^6\,\rm T/m$ \cite{Rugar2004}.
Finally our findings demonstrate how MW dressing rotates the perspective in the Bloch sphere dynamics and transforms a parametric interaction into a resonant coupling between the mechanical oscillator and the dressed states, so that one can expect the wealth of cavity QED experimental signatures \cite{Haroche2006,Sanchez-Mondragon1983} to be transposed to qubit-mechanical hybrid systems in parametric interaction.\\
\textit{Acknowledgements---} We thank  C. Fabre, O. Buisson, A. Auffeves, G. Nogues, D. Feinberg, S. Seidelin, M. Dartiailh, C. Hoarau and D. Lepoittevin, for theoretical, experimental and technical assistance. This project is supported by a Marie Curie reintegration grant,  an ANR RPDoc program and the ERC Starting Grant HQ-NOM. S.R. acknowledges funding from the Nanoscience Foundation.


\begin{thebibliography}{0}
\expandafter\ifx\csname natexlab\endcsname\relax\def\natexlab#1{#1}\fi
\expandafter\ifx\csname bibnamefont\endcsname\relax
  \def\bibnamefont#1{#1}\fi
\expandafter\ifx\csname bibfnamefont\endcsname\relax
  \def\bibfnamefont#1{#1}\fi
\expandafter\ifx\csname citenamefont\endcsname\relax
  \def\citenamefont#1{#1}\fi
\expandafter\ifx\csname url\endcsname\relax
  \def\url#1{\texttt{#1}}\fi
\expandafter\ifx\csname urlprefix\endcsname\relax\def\urlprefix{URL }\fi
\providecommand{\bibinfo}[2]{#2}
\providecommand{\eprint}[2][]{\url{#2}}

\end{thebibliography}


\begin{thebibliography}{46}
\expandafter\ifx\csname natexlab\endcsname\relax\def\natexlab#1{#1}\fi
\expandafter\ifx\csname bibnamefont\endcsname\relax
  \def\bibnamefont#1{#1}\fi
\expandafter\ifx\csname bibfnamefont\endcsname\relax
  \def\bibfnamefont#1{#1}\fi
\expandafter\ifx\csname citenamefont\endcsname\relax
  \def\citenamefont#1{#1}\fi
\expandafter\ifx\csname url\endcsname\relax
  \def\url#1{\texttt{#1}}\fi
\expandafter\ifx\csname urlprefix\endcsname\relax\def\urlprefix{URL }\fi
\providecommand{\bibinfo}[2]{#2}
\providecommand{\eprint}[2][]{\url{#2}}

\bibitem[{\citenamefont{Schwab and Roukes}(2005)}]{Schwab2005}
\bibinfo{author}{\bibfnamefont{K.~C.} \bibnamefont{Schwab}} \bibnamefont{and}
  \bibinfo{author}{\bibfnamefont{M.~L.} \bibnamefont{Roukes}},
  \bibinfo{journal}{Phys. Today} \textbf{\bibinfo{volume}{58}}
  (\bibinfo{year}{2005}).

\bibitem[{\citenamefont{Treutlein~{\it et al.}}(2013)}]{Treutlein2013}
\bibinfo{author}{\bibfnamefont{P.}~\bibnamefont{Treutlein~{\it et al.}}},
  \bibinfo{journal}{arXiv:1210.4151}  (\bibinfo{year}{2013}).

\bibitem[{\citenamefont{LaHaye~{\it et al.}}(2009)}]{LaHaye2009}
\bibinfo{author}{\bibfnamefont{M.~D.} \bibnamefont{LaHaye~{\it et al.}}},
  \bibinfo{journal}{Nature} \textbf{\bibinfo{volume}{459}},
  \bibinfo{pages}{960} (\bibinfo{year}{2009}).

\bibitem[{\citenamefont{Pirkkalainen~{\it et al.}}(2013)}]{Pirkkalainen2013}
\bibinfo{author}{\bibfnamefont{J.-M.} \bibnamefont{Pirkkalainen~{\it et al.}}},
  \bibinfo{journal}{Nature} \textbf{\bibinfo{volume}{494}},
  \bibinfo{pages}{211} (\bibinfo{year}{2013}).

\bibitem[{\citenamefont{Treutlein et~al.}(2007)\citenamefont{Treutlein, Hunger,
  Camerer, Hansch, and Reichel}}]{Treutlein2007}
\bibinfo{author}{\bibfnamefont{P.}~\bibnamefont{Treutlein}},
  \bibinfo{author}{\bibfnamefont{D.}~\bibnamefont{Hunger}},
  \bibinfo{author}{\bibfnamefont{S.}~\bibnamefont{Camerer}},
  \bibinfo{author}{\bibfnamefont{T.}~\bibnamefont{Hansch}}, \bibnamefont{and}
  \bibinfo{author}{\bibfnamefont{J.}~\bibnamefont{Reichel}},
  \bibinfo{journal}{Physical Review Letters} \textbf{\bibinfo{volume}{99}},
  \bibinfo{pages}{140403} (\bibinfo{year}{2007}).

\bibitem[{\citenamefont{Lassagne~{\it et al.}}(2009)}]{Lassagne2009}
\bibinfo{author}{\bibfnamefont{B.}~\bibnamefont{Lassagne~{\it et al.}}},
  \bibinfo{journal}{Science} \textbf{\bibinfo{volume}{325}},
  \bibinfo{pages}{1107} (\bibinfo{year}{2009}).

\bibitem[{\citenamefont{Steele{\it et al.}}(2009)}]{Steele2009}
\bibinfo{author}{\bibfnamefont{G.}~\bibnamefont{Steele{\it et al.}}},
  \bibinfo{journal}{Science} \textbf{\bibinfo{volume}{325}},
  \bibinfo{pages}{1103} (\bibinfo{year}{2009}).

\bibitem[{\citenamefont{Sallen~{\it et al.}}(2009)}]{Sallen2009}
\bibinfo{author}{\bibfnamefont{G.}~\bibnamefont{Sallen~{\it et al.}}},
  \bibinfo{journal}{Physical Review B} \textbf{\bibinfo{volume}{80}},
  \bibinfo{pages}{085310} (\bibinfo{year}{2009}).

\bibitem[{\citenamefont{Bennett et~al.}(2010)\citenamefont{Bennett, Cockins,
  Miyahara, Grutter, and Clerk}}]{Bennett2010}
\bibinfo{author}{\bibfnamefont{S.}~\bibnamefont{Bennett}},
  \bibinfo{author}{\bibfnamefont{L.}~\bibnamefont{Cockins}},
  \bibinfo{author}{\bibfnamefont{Y.}~\bibnamefont{Miyahara}},
  \bibinfo{author}{\bibfnamefont{P.}~\bibnamefont{Grutter}}, \bibnamefont{and}
  \bibinfo{author}{\bibfnamefont{A.}~\bibnamefont{Clerk}},
  \bibinfo{journal}{Phys. Rev. Lett.} \textbf{\bibinfo{volume}{104}},
  \bibinfo{pages}{017203} (\bibinfo{year}{2010}).

\bibitem[{\citenamefont{Ganzhorn~{\it et al.}}(2013)}]{Ganzhorn2013}
\bibinfo{author}{\bibfnamefont{M.}~\bibnamefont{Ganzhorn~{\it et al.}}},
  \bibinfo{journal}{Nature Nanotech.} \textbf{\bibinfo{volume}{8}},
  \bibinfo{pages}{165} (\bibinfo{year}{2013}).

\bibitem[{\citenamefont{Blatt and Wineland}(2008)}]{Blatt2008}
\bibinfo{author}{\bibfnamefont{R.}~\bibnamefont{Blatt}} \bibnamefont{and}
  \bibinfo{author}{\bibfnamefont{D.~J.} \bibnamefont{Wineland}},
  \bibinfo{journal}{Nature} \textbf{\bibinfo{volume}{453}},
  \bibinfo{pages}{1008} (\bibinfo{year}{2008}).

\bibitem[{\citenamefont{Hanson et~al.}(2006)\citenamefont{Hanson, Gywat, and
  Awschalom}}]{Hanson2006}
\bibinfo{author}{\bibfnamefont{R.}~\bibnamefont{Hanson}},
  \bibinfo{author}{\bibfnamefont{O.}~\bibnamefont{Gywat}}, \bibnamefont{and}
  \bibinfo{author}{\bibfnamefont{D.}~\bibnamefont{Awschalom}},
  \bibinfo{journal}{Phys. Rev. B} \textbf{\bibinfo{volume}{74}},
  \bibinfo{pages}{161203} (\bibinfo{year}{2006}).

\bibitem[{\citenamefont{Childress~{\it et al.}}(2006)}]{Childress2006}
\bibinfo{author}{\bibfnamefont{L.}~\bibnamefont{Childress~{\it et al.}}},
  \bibinfo{journal}{Science} \textbf{\bibinfo{volume}{314}},
  \bibinfo{pages}{281} (\bibinfo{year}{2006}).

\bibitem[{\citenamefont{Balasubramanian~{\it et
  al.}}(2009)}]{Balasubramanian2009}
\bibinfo{author}{\bibfnamefont{G.}~\bibnamefont{Balasubramanian~{\it et al.}}},
  \bibinfo{journal}{Nature Mat.} \textbf{\bibinfo{volume}{8}},
  \bibinfo{pages}{383} (\bibinfo{year}{2009}).

\bibitem[{\citenamefont{Rabl~{\it et al.}}(2009)}]{Rabl2009}
\bibinfo{author}{\bibfnamefont{P.}~\bibnamefont{Rabl~{\it et al.}}},
  \bibinfo{journal}{Physical Review B} \textbf{\bibinfo{volume}{79}},
  \bibinfo{pages}{041302} (\bibinfo{year}{2009}).

\bibitem[{\citenamefont{Rabl}(2010)}]{Rabl2010}
\bibinfo{author}{\bibfnamefont{P.}~\bibnamefont{Rabl}},
  \bibinfo{journal}{Physical Review B} \textbf{\bibinfo{volume}{82}},
  \bibinfo{pages}{165320} (\bibinfo{year}{2010}).

\bibitem[{\citenamefont{Arcizet~{\it et al.}}(2011)}]{Arcizet2011a}
\bibinfo{author}{\bibfnamefont{O.}~\bibnamefont{Arcizet~{\it et al.}}},
  \bibinfo{journal}{Nature Physics} \textbf{\bibinfo{volume}{7}},
  \bibinfo{pages}{879} (\bibinfo{year}{2011}).

\bibitem[{\citenamefont{Bennett~{\it et al.}}(2012)}]{Bennett2012}
\bibinfo{author}{\bibfnamefont{S.~D.} \bibnamefont{Bennett~{\it et al.}}},
  \bibinfo{journal}{New Journal of Physics} \textbf{\bibinfo{volume}{14}},
  \bibinfo{pages}{125004} (\bibinfo{year}{2012}).

\bibitem[{\citenamefont{Hong~{\it et al.}}(2012)}]{Yacoby2012}
\bibinfo{author}{\bibfnamefont{S.}~\bibnamefont{Hong~{\it et al.}}},
  \bibinfo{journal}{Nano letters} \textbf{\bibinfo{volume}{12}},
  \bibinfo{pages}{3920} (\bibinfo{year}{2012}).

\bibitem[{\citenamefont{Rugar~{\it et al.}}(2004)}]{Rugar2004}
\bibinfo{author}{\bibfnamefont{D.}~\bibnamefont{Rugar~{\it et al.}}},
  \bibinfo{journal}{Nature} \textbf{\bibinfo{volume}{430}},
  \bibinfo{pages}{329} (\bibinfo{year}{2004}).

\bibitem[{\citenamefont{Degen~{\it et al.}}(2009)}]{Degen2009}
\bibinfo{author}{\bibfnamefont{C.~L.} \bibnamefont{Degen~{\it et al.}}},
  \bibinfo{journal}{Proc. Nat. Acad. Sc.} \textbf{\bibinfo{volume}{106}},
  \bibinfo{pages}{1313} (\bibinfo{year}{2009}).

\bibitem[{\citenamefont{Nichol et~al.}(2012)\citenamefont{Nichol, Hemesath,
  Lauhon, and Budakian}}]{Nichol2012}
\bibinfo{author}{\bibfnamefont{J.}~\bibnamefont{Nichol}},
  \bibinfo{author}{\bibfnamefont{E.}~\bibnamefont{Hemesath}},
  \bibinfo{author}{\bibfnamefont{L.}~\bibnamefont{Lauhon}}, \bibnamefont{and}
  \bibinfo{author}{\bibfnamefont{R.}~\bibnamefont{Budakian}},
  \bibinfo{journal}{Physical Review B} \textbf{\bibinfo{volume}{85}},
  \bibinfo{pages}{054414} (\bibinfo{year}{2012}).

\bibitem[{\citenamefont{Childress and McIntyre}(2010)}]{Childress2010}
\bibinfo{author}{\bibfnamefont{L.}~\bibnamefont{Childress}} \bibnamefont{and}
  \bibinfo{author}{\bibfnamefont{J.}~\bibnamefont{McIntyre}},
  \bibinfo{journal}{Phys. Rev. A} \textbf{\bibinfo{volume}{82}},
  \bibinfo{pages}{033839} (\bibinfo{year}{2010}).

\bibitem[{\citenamefont{Li~{\it et al.}}(2013)}]{Li2013}
\bibinfo{author}{\bibfnamefont{J.}~\bibnamefont{Li~{\it et al.}}},
  \bibinfo{journal}{Nature communications} \textbf{\bibinfo{volume}{4}},
  \bibinfo{pages}{1420} (\bibinfo{year}{2013}).

\bibitem[{\citenamefont{Jelezko et~al.}(2004)\citenamefont{Jelezko, Gaebel,
  Popa, Gruber, and Wrachtrup}}]{Jelezko2004}
\bibinfo{author}{\bibfnamefont{F.}~\bibnamefont{Jelezko}},
  \bibinfo{author}{\bibfnamefont{T.}~\bibnamefont{Gaebel}},
  \bibinfo{author}{\bibfnamefont{I.}~\bibnamefont{Popa}},
  \bibinfo{author}{\bibfnamefont{A.}~\bibnamefont{Gruber}}, \bibnamefont{and}
  \bibinfo{author}{\bibfnamefont{J.}~\bibnamefont{Wrachtrup}},
  \bibinfo{journal}{Phys. Rev. Lett.} \textbf{\bibinfo{volume}{92}},
  \bibinfo{pages}{076401} (\bibinfo{year}{2004}).

\bibitem[{\citenamefont{Slichter}(1996)}]{Slichter1996}
\bibinfo{author}{\bibfnamefont{C.~P.} \bibnamefont{Slichter}},
  \emph{\bibinfo{title}{{Principles of Magnetic Resonance}}}
  (\bibinfo{publisher}{Springer}, \bibinfo{address}{Heidelberg},
  \bibinfo{year}{1996}).

\bibitem[{\citenamefont{Holmstrom et~al.}(1997)\citenamefont{Holmstrom, Wei,
  Windsor, Manson, Martin, and Glasbeek}}]{Holmstrom1997}
\bibinfo{author}{\bibfnamefont{S.}~\bibnamefont{Holmstrom}},
  \bibinfo{author}{\bibfnamefont{C.}~\bibnamefont{Wei}},
  \bibinfo{author}{\bibfnamefont{A.}~\bibnamefont{Windsor}},
  \bibinfo{author}{\bibfnamefont{N.}~\bibnamefont{Manson}},
  \bibinfo{author}{\bibfnamefont{J.}~\bibnamefont{Martin}}, \bibnamefont{and}
  \bibinfo{author}{\bibfnamefont{M.}~\bibnamefont{Glasbeek}},
  \bibinfo{journal}{Phys. Rev. Lett.} \textbf{\bibinfo{volume}{78}},
  \bibinfo{pages}{302} (\bibinfo{year}{1997}).

\bibitem[{\citenamefont{Holmstrom et~al.}(1998)\citenamefont{Holmstrom,
  Windsor, Wei, Martin, and Manson}}]{Holmstrom1998}
\bibinfo{author}{\bibfnamefont{S.~A.} \bibnamefont{Holmstrom}},
  \bibinfo{author}{\bibfnamefont{A.~S.~M.} \bibnamefont{Windsor}},
  \bibinfo{author}{\bibfnamefont{C.}~\bibnamefont{Wei}},
  \bibinfo{author}{\bibfnamefont{J.~P.~D.} \bibnamefont{Martin}},
  \bibnamefont{and} \bibinfo{author}{\bibfnamefont{N.~B.}
  \bibnamefont{Manson}}, \bibinfo{journal}{Journal of Luminescence}
  \textbf{\bibinfo{volume}{76}}, \bibinfo{pages}{38} (\bibinfo{year}{1998}).

\bibitem[{\citenamefont{Saiko and Fedoruk}(2008)}]{Saiko2008}
\bibinfo{author}{\bibfnamefont{A.}~\bibnamefont{Saiko}} \bibnamefont{and}
  \bibinfo{author}{\bibfnamefont{G.}~\bibnamefont{Fedoruk}},
  \bibinfo{journal}{JETP Letters} \textbf{\bibinfo{volume}{3}},
  \bibinfo{pages}{128} (\bibinfo{year}{2008}).

\bibitem[{\citenamefont{Saiko and Fedaruk}(2010)}]{Saiko2010}
\bibinfo{author}{\bibfnamefont{A.~P.} \bibnamefont{Saiko}} \bibnamefont{and}
  \bibinfo{author}{\bibfnamefont{R.}~\bibnamefont{Fedaruk}},
  \bibinfo{journal}{JETP Letters} \textbf{\bibinfo{volume}{91}},
  \bibinfo{pages}{681} (\bibinfo{year}{2010}).

\bibitem[{\citenamefont{Kotler et~al.}(2011)\citenamefont{Kotler, Akerman,
  Glickman, Keselman, and Ozeri}}]{Kotler2011}
\bibinfo{author}{\bibfnamefont{S.}~\bibnamefont{Kotler}},
  \bibinfo{author}{\bibfnamefont{N.}~\bibnamefont{Akerman}},
  \bibinfo{author}{\bibfnamefont{Y.}~\bibnamefont{Glickman}},
  \bibinfo{author}{\bibfnamefont{A.}~\bibnamefont{Keselman}}, \bibnamefont{and}
  \bibinfo{author}{\bibfnamefont{R.}~\bibnamefont{Ozeri}},
  \bibinfo{journal}{Nature} \textbf{\bibinfo{volume}{473}}, \bibinfo{pages}{61}
  (\bibinfo{year}{2011}).

\bibitem[{\citenamefont{deLange et~al.}(2011)\citenamefont{deLange, Riste,
  Dobrovitski, and Hanson}}]{Lange2011}
\bibinfo{author}{\bibfnamefont{G.}~\bibnamefont{deLange}},
  \bibinfo{author}{\bibfnamefont{D.}~\bibnamefont{Riste}},
  \bibinfo{author}{\bibfnamefont{V.}~\bibnamefont{Dobrovitski}},
  \bibnamefont{and} \bibinfo{author}{\bibfnamefont{R.}~\bibnamefont{Hanson}},
  \bibinfo{journal}{Phys. Rev. Lett.} \textbf{\bibinfo{volume}{106}},
  \bibinfo{pages}{080802} (\bibinfo{year}{2011}).

\bibitem[{\citenamefont{Loretz et~al.}(2013)\citenamefont{Loretz, Rosskopf, and
  Degen}}]{Degen2013}
\bibinfo{author}{\bibfnamefont{M.}~\bibnamefont{Loretz}},
  \bibinfo{author}{\bibfnamefont{T.}~\bibnamefont{Rosskopf}}, \bibnamefont{and}
  \bibinfo{author}{\bibfnamefont{C.}~\bibnamefont{Degen}},
  \bibinfo{journal}{Phys. Rev. Lett.} \textbf{\bibinfo{volume}{110}},
  \bibinfo{pages}{017602} (\bibinfo{year}{2013}).

\bibitem[{\citenamefont{Haroche and Raimond}(2006)}]{Haroche2006}
\bibinfo{author}{\bibfnamefont{S.}~\bibnamefont{Haroche}} \bibnamefont{and}
  \bibinfo{author}{\bibfnamefont{J.~M.} \bibnamefont{Raimond}},
  \emph{\bibinfo{title}{Exploring the quantum}} (\bibinfo{publisher}{Oxford
  University Press}, \bibinfo{year}{2006}).

\bibitem[{\citenamefont{Kolkowitz~{\it et al.}}(2012)}]{Kolkowitz2012}
\bibinfo{author}{\bibfnamefont{S.}~\bibnamefont{Kolkowitz~{\it et al.}}},
  \bibinfo{journal}{Science} \textbf{\bibinfo{volume}{335}},
  \bibinfo{pages}{1603} (\bibinfo{year}{2012}).

\bibitem[{\citenamefont{Puller~{\it et al.}}(2013)}]{Puller2013}
\bibinfo{author}{\bibfnamefont{V.}~\bibnamefont{Puller~{\it et al.}}},
  \bibinfo{journal}{Phys. Rev. Lett.} \textbf{\bibinfo{volume}{110}},
  \bibinfo{pages}{125501} (\bibinfo{year}{2013}).

\bibitem[{\citenamefont{Oliver~{\it et al.}}(2005)}]{Oliver2005}
\bibinfo{author}{\bibfnamefont{W.~D.} \bibnamefont{Oliver~{\it et al.}}},
  \bibinfo{journal}{Science} \textbf{\bibinfo{volume}{310}},
  \bibinfo{pages}{1653} (\bibinfo{year}{2005}).

\bibitem[{\citenamefont{Dupont-Ferrier~{\it et al.}}(2013)}]{DupontFerrier2013}
\bibinfo{author}{\bibfnamefont{E.}~\bibnamefont{Dupont-Ferrier~{\it et al.}}},
  \bibinfo{journal}{Phys. Rev. Lett.} \textbf{\bibinfo{volume}{110}},
  \bibinfo{pages}{136802} (\bibinfo{year}{2013}).

\bibitem[{\citenamefont{Dreau et~al.}(2011)\citenamefont{Dreau, Lesik, Rondin,
  Spinicelli, Arcizet, Roch, and Jacques}}]{Dreau2011}
\bibinfo{author}{\bibfnamefont{A.}~\bibnamefont{Dreau}},
  \bibinfo{author}{\bibfnamefont{M.}~\bibnamefont{Lesik}},
  \bibinfo{author}{\bibfnamefont{L.}~\bibnamefont{Rondin}},
  \bibinfo{author}{\bibfnamefont{P.}~\bibnamefont{Spinicelli}},
  \bibinfo{author}{\bibfnamefont{O.}~\bibnamefont{Arcizet}},
  \bibinfo{author}{\bibfnamefont{J.-F.} \bibnamefont{Roch}}, \bibnamefont{and}
  \bibinfo{author}{\bibfnamefont{V.}~\bibnamefont{Jacques}},
  \bibinfo{journal}{Physical Review B} \textbf{\bibinfo{volume}{84}},
  \bibinfo{pages}{195204} (\bibinfo{year}{2011}).

\bibitem[{\citenamefont{Leibfried~{\it et al.}}(2003)}]{Leibfried2003}
\bibinfo{author}{\bibfnamefont{D.}~\bibnamefont{Leibfried~{\it et al.}}},
  \bibinfo{journal}{Nature} \textbf{\bibinfo{volume}{422}},
  \bibinfo{pages}{412} (\bibinfo{year}{2003}).

\bibitem[{\citenamefont{Schliesser~{\it et al.}}(2009)}]{Schliesser2009}
\bibinfo{author}{\bibfnamefont{A.}~\bibnamefont{Schliesser~{\it et al.}}},
  \bibinfo{journal}{Nature Physics} \textbf{\bibinfo{volume}{5}},
  \bibinfo{pages}{509} (\bibinfo{year}{2009}).

\bibitem[{\citenamefont{Jacques~{\it et al.}}(2009)}]{Jacques2009}
\bibinfo{author}{\bibfnamefont{V.}~\bibnamefont{Jacques~{\it et al.}}},
  \bibinfo{journal}{Phys. Rev. Lett.} \textbf{\bibinfo{volume}{102}},
  \bibinfo{pages}{057403} (\bibinfo{year}{2009}).

\bibitem[{\citenamefont{Smeltzer et~al.}(2009)\citenamefont{Smeltzer, McIntyre,
  and Childress}}]{Smeltzer2009}
\bibinfo{author}{\bibfnamefont{B.}~\bibnamefont{Smeltzer}},
  \bibinfo{author}{\bibfnamefont{J.}~\bibnamefont{McIntyre}}, \bibnamefont{and}
  \bibinfo{author}{\bibfnamefont{L.}~\bibnamefont{Childress}},
  \bibinfo{journal}{Physical Review A} \textbf{\bibinfo{volume}{80}},
  \bibinfo{pages}{050302} (\bibinfo{year}{2009}).

\bibitem[{\citenamefont{Autler and Townes}(1955)}]{Autler1955}
\bibinfo{author}{\bibfnamefont{S.}~\bibnamefont{Autler}} \bibnamefont{and}
  \bibinfo{author}{\bibfnamefont{C.}~\bibnamefont{Townes}},
  \bibinfo{journal}{Phys. Rev.} \textbf{\bibinfo{volume}{100}},
  \bibinfo{pages}{703} (\bibinfo{year}{1955}).

\bibitem[{\citenamefont{Cohen-Tannoudji~{\it et
  al.}}(1973)}]{Cohen-Tannoudji1973}
\bibinfo{author}{\bibfnamefont{C.}~\bibnamefont{Cohen-Tannoudji~{\it et al.}}},
  \bibinfo{journal}{J. Phys. B: Atom. Molec. Phys.}
  \textbf{\bibinfo{volume}{6}}, \bibinfo{pages}{214} (\bibinfo{year}{1973}).

\bibitem[{\citenamefont{Sanchez-Mondragon
  et~al.}(1983)\citenamefont{Sanchez-Mondragon, Narozhny, and
  Eberly}}]{Sanchez-Mondragon1983}
\bibinfo{author}{\bibfnamefont{J.}~\bibnamefont{Sanchez-Mondragon}},
  \bibinfo{author}{\bibfnamefont{N.}~\bibnamefont{Narozhny}}, \bibnamefont{and}
  \bibinfo{author}{\bibfnamefont{J.}~\bibnamefont{Eberly}},
  \bibinfo{journal}{Phys. Rev. Lett.} \textbf{\bibinfo{volume}{51}},
  \bibinfo{pages}{550} (\bibinfo{year}{1983}).

\end{thebibliography}
\end{document}